\documentclass[preprint,showpacs,preprintnumbers,amsmath,amssymb,aps]{revtex4}

\usepackage{graphicx}
\usepackage{dcolumn}
\usepackage{bm}

\begin {document}
\title{\Large\bf
Transverse single spin asymmetry in Drell-Yan production in
polarized pA collisions }

\author{Jian Zhou
 \\[0.3cm]
{\normalsize\it Nikhef and Department of Physics and Astronomy, VU University Amsterdam,
 De Boelelaan 1081, NL-1081 HV Amsterdam, the Netherlands
}}

\begin{abstract}
\noindent
 We study the transverse single spin asymmetry in Drell-Yan
  production in pA collisions with incoming protons being
transversely polarized. We carry out the calculation using a newly developed hybrid approach.
The polarized cross section computed in the hybrid approach is consistent with that obtained from the usual TMD
factorization at low transverse momentum as expected, whereas  at high transverse momentum,
color entanglement effect is found to play a role in contributing to the
spin asymmetry of Drell-Yan production, though it is a $1/N_c^2$ suppressed effect.
\end{abstract}

\maketitle

\section{Introduction}
Proton-nucleus(or deuteron-nucleus) collisions at RHIC  provide an unique opportunity to study
saturation/Color Class Condensate(CGC) physics. Many relevant observables in
pA collisions in the forward rapidities region are excellent probes for accessing the saturated
small $x$ gluon distributions inside nucleus.
The remarkable theoretical and experimental progress made recently in the field mostly focus on the
spin independent observables, among which the single hadron suppression and the di-hadron correlation at forward
rapidities play important roles in studying saturation physics~\cite{Arsene:2004ux}. Meanwhile, polarized proton-proton collisions
at RHIC have made a big impact on the investigation of the nucleon's spin structure.
In particular, transverse single spin asymmetries(SSAs) phenomena in polarized pp collisions
 have gained a lot of attentions~\cite{Adams:2003fx},
 as the study of SSAs not only could help us to map out the three dimensional image of nucleon~\cite{Boer:2011fh},
 but also  greatly deepened our understanding of QCD and its associated factorization properties.

On the other hand, scattering a polarized probe on a dense background gluon field inside a large nucleus may provide a promising way of studying
the interplay of saturation effects and transverse spin phenomena.
The authors of paper~\cite{Kang:2011ni} have proposed to probe the saturation scale of nucleus by measuring SSAs normalized by that in
pp scattering at low transverse momentum.
It is also important to measure the SSA for prompt photon production in polarized pA collisions in order to
distinguish different mechanisms for generating the SSA~\cite{Kovchegov:2012ga,Schafer:2014zea,Kanazawa:2014nea}.
Furthermore, it has been shown that polarized observables are sensitive to
the slope  of small $x$ gluon transverse momentum dependent(TMD) distributions in $k_\perp$ space~\cite{Boer:2002ij,Boer:2006rj,Schafer:2014xpa}.
Measuring transverse momentum dependence of SSAs thus may provide complementary information on small $x$ gluon TMDs.

 Polarized pA collisions also present an advantage over unpolarized pA collisions
and polarized pp collisions in addressing one novel aspect of QCD: color entanglement effect~\cite{Rogers:2010dm}.
To generate the imaginary phase necessary for the non-vanishing SSAs, one additional gluon must be exchanged between the
partonic hard part and the   proton remnant.
The interactions of this additional gluon and the valence quark from proton with
the saturated gluon field inside nucleus lead to a very complicate color flow structure which could
 give rise to  color entanglement.
 Such effect is the consequence of nontrivial interplay among the T-odd effect, the coherent multiple
 gluon re-scattering, and the non-Abelian feature of QCD.
 Investigating SSAs in polarized pA collisions may shed new
 light on the study of generalized TMD factorization breaking effect that is caused by color entanglement.
A polarized pA collisions program at RHIC is therefore extremely welcome~\cite{Aschenauer:2013woa}.

In this paper, we study the SSA in Drell-Yan lepton pair production at forward rapidities in polarized pA collisions.
Due to the absence of final state interactions and fragmentation effects, the SSA in Drell-Yan process offers a
very clean probe for the Sivers effect.
The contribution from the Sivers effect to the SSAs has been well
formulated in the context of TMD factorization and the collinear twist-3 approach.
For the polarized pA collisions case, to incorporate the saturation effect,
 we carry out the calculation in a hybrid approach in which  the nucleus is treated in the
 CGC framework while the collinear twist-3 formalism is applied on the proton side.
Such hybrid approach has been recently developed to study the SSAs in prompt photon production and
photon-jet production in polarized pA collisions~\cite{Schafer:2014zea,Schafer:2014xpa}.
We notice that the SSA in Drell-Yan production has also been studied using a different hybrid approach~\cite{Kang:2012vm}.
As shown below,  two different hybrid approaches yield the same result for this observable
 at low transverse momentum.

In a more general context, the present work is part of the effort to address
 the interplay between spin physics and saturation physics.
Apart from the studies mentioned above, recent work in this very
active field includes the study of  the
quark Boer-Mulders distribution and the linearly polarized gluon
distribution inside a large nucleus~\cite{Metz:2011wb}.
The small $x$ evolution equations for the linearly polarized gluon
distributions were derived in Ref.~\cite{Dominguez:2011br}.
The first numerical study of the linearly polarized gluon distribution was presented in~\cite{Dumitru:2014vka}.
Furthermore, the asymptotic behavior of transverse single spin
 asymmetries at small $x$ was discussed in Ref.~\cite{Schafer:2013opa,Zhou:2013gsa}.
It has been shown that SSAs at small $x$ are generated by the spin dependent
odderon exchange whose size is determined by the anomalous magnetic
moment of proton~\cite{Zhou:2013gsa}. The quark Sivers function was computed
in the  Glauber-Mueller/McLerran-Venugopalan(MV) models~\cite{Kovchegov:2013cva}.
The spin asymmetries in pA collisions have been investigated by going beyond the Eikonal approximation within
the CGC framework~\cite{Altinoluk:2014oxa}.

The paper is structured as follows. In section II, we derive the spin dependent amplitude using the hybrid approach,
including both soft gluon pole and hard gluon pole contributions. In section III, we present  expressions
for the polarized cross section in different kinematic limits and compare our results with
 that obtained from different approaches which are applicable in the corresponding kinematic regions.
 The paper is summarized in section IV.

\section{The derivation of the spin dependent amplitude  }
 In this section, we derive the spin dependent amplitude for Drell-Yan production using the newly developed
 hybrid approach.  We start by briefly reviewing the CGC calculation for unpolarized Drell-Yan cross section
 in pA collisions.

The dominant production mechanism for Drell-Yan virtual photons at forward rapidities
 is Compton scattering $qg \rightarrow \gamma^* q$.
We fix the relevant kinematical variables and assign 4-momenta to the particles according to
\begin{eqnarray}
\ q(xP)\ + \  g(x_g' \bar P+k_\perp ) \longrightarrow \gamma^*(l_{\gamma^*}) \  + \ q(l_q)
\end{eqnarray}
where $\bar P^\mu=\bar P^- n^\mu$ and  $ P^\mu= P^+ p^\mu$ with
$n^\mu$ and $p^\mu$ being the commonly defined light cone vectors,
normalized according to $p \cdot n=1$. The Mandelstam variables are
defined as: $S=(P+\bar P)^2$, $T=(P-l_q)^2$ and
$U=(P-l_{\gamma^*})^2$. The invariant mass of the produced lepton pair is denoted as $M^2=l_{\gamma^*}^2$.
It is worthy to mention that $x_g' \bar P+k_\perp$ is the total momentum transfer via multiple gluon re-scattering.

The calculation for Drell-Yan virtual photon production in unpolarized pA collisions
is rather similar to that for prompt photon production,  and has been done within the CGC framework a decade ago~\cite{Gelis:2002fw}.
The key ingredient of this calculation is resumming multiple gluon re-scattering into  Wilson line
 which is a path-ordered gauge factor along the straight line that extends in
$x^+$ from minus infinity to plus infinity.
More precisely, for a quark with incoming momentum
 $l$ and outgoing momentum  $l+k$, the path-ordered gauge factor reads,
\begin{equation}
 2 \pi  \delta(k^+)
 n^\mu [U(k_\perp)-(2\pi)^2 \delta(k_\perp)] \,,
\end{equation}
with
\begin{equation}
U(k_\perp) =\int d^2 x_\perp e^{ i  k_\perp \cdot  x_\perp}
U(x_\perp)\,,
\end{equation}
and
\begin{equation}
U(x_\perp)= {\cal P} e^{ig\int_{-\infty}^{+\infty} dx^+ A^-_A (x^+,
\ x_\perp)\cdot t }  \ ,
\end{equation}
where  $t$ is the generators in the fundamental representation.
With this calculation recipe, it is straightforward to obtain
 the cross section for unpolarized Drell-Yan lepton pair  production~\cite{Gelis:2002fw},
 \begin{eqnarray}
\frac{d^3  \sigma}{dM^2 d^2 l_{\gamma^*\perp} dy }=
 \frac{ \alpha_{em}^2 \alpha_s}{3\pi M^2 N_c}
 \sum_q e_q^2  \int \frac{dx}{x}  d^2 k_\perp \
H(k_\perp, l_{\gamma^* \perp},z)   xf_q(x)  x_g'G_{DP}(x_g',k_\perp)   \label{unp}
\end{eqnarray}
where $y$ is the rapidity of the virtual photon. In the above formula,
 $f_q(x)$ is the integrated unpolarized quark distribution from proton,
  and $x_g'G_{DP}(x_g',k_\perp)$ is the dipole type gluon TMD, defined as,
\begin{eqnarray}
x_g' G_{DP}(x_g', k_\perp)=\frac{k_\perp^2 N_c}{2 \pi^2 \alpha_s}
\int \frac{d^2x_\perp d^2y_\perp }{(2\pi)^2} e^{i  k_\perp \cdot
(y_\perp-x_\perp)} \frac{1}{N_c} \langle {\rm Tr} \left [
U(x_\perp)U^\dag(y_\perp)  \right ] \rangle_{x_g'}
\end{eqnarray}
The hard part $H(k_\perp, l_{\gamma^* \perp},z)$ is given by,
 \begin{eqnarray}
 H(k_\perp, l_{\gamma^* \perp},z)&=&[1+(1-z)^2]
 \frac{z^2 } {\left [ (l_{\gamma^* \perp}-zk_\perp)^2+\epsilon^2_M \right ]\left [l_{\gamma^* \perp}^2+\epsilon^2_M \right ]}
\nonumber \\
&&-z^2 \epsilon^2_M \frac{1}{k_\perp^2} \left [ \frac{1}{l_{\gamma^* \perp}^2+\epsilon^2_M}
-\frac{1}{ (l_{\gamma^* \perp}-zk_\perp)^2+\epsilon^2_M } \right ]^2
 \end{eqnarray}
where $z\equiv l_{\gamma^*} \cdot n/(x P \cdot n) $ is the fraction of
the incoming quark momentum $xP$ carried by the virtual photon,
 and $\epsilon^2_M=(1-z)M^2$.
 $l_{\gamma^*\perp}$ is the virtual photon transverse momentum.
 The connections between the above Drell-Yan cross section derived in the CGC framework and
 those from the TMD factorization and the collinear factorization have been discussed in paper~\cite{Kang:2012vm}.

\begin{figure}[t]
\begin{center}
\includegraphics[width=11 cm]{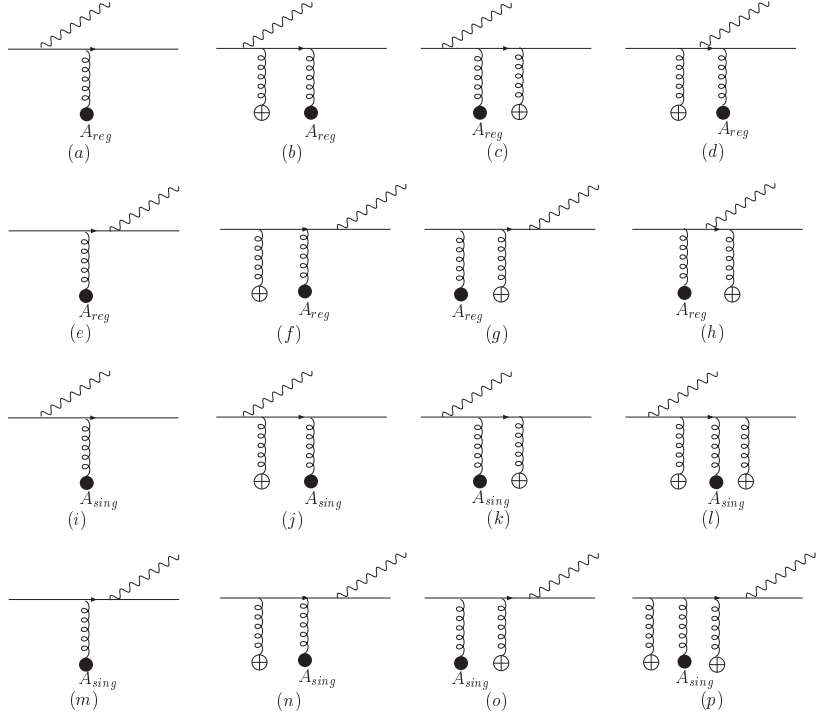}
\caption[] { Diagrams contributing to the spin dependent Drell-Yan lepton pair production amplitude.
Different symbols indicate different parts of the classical gluon field.
A black dot denotes $A_{reg}$ or $A_{sing}$, while  a cross
surrounded with a circle denotes $A_A$. Diagrams Fig.1a, Fig.1b, Fig.1d, Fig.1e, Fig.1f, and Fig.1i-Fig.1p generate
the soft gluon pole contribution.
Diagrams Fig.1a-Fig.1d and Fig.1i-Fig.1l give rise to the
hard gluon pole contribution.}
 \label{1}
\end{center}
\end{figure}

We now move on to derive the spin dependent amplitude of Drell-Yan production.
To generate the spin asymmetry, one additional gluon must be exchanged between the active partons
and the remnant part of the polarized proton projectile.
In the collinear twist-3 approach, the associated soft part is described by
the three parton correlator: the ETQS function~\cite{Efremov:1981sh,Qiu:1991pp},
\begin{eqnarray}
T_{F,q}(x_1,x)&=&\int \frac{dy_1^-dy_2^-}{4 \pi} e^{ix_1P^+ y_1^-
+ i(x-x_1)P^+y^-}
\nonumber \\
&& \times \langle P,S_\perp | \bar{\psi}_q(0)\gamma^+ g
\epsilon^{S_\perp \sigma n p} F_ \sigma^{ \ +}(y^-)\psi_q(y_1^-) |
P,S_\perp \rangle
\end{eqnarray}
where we have suppressed Wilson lines. $S_\perp$ denotes the proton
transverse spin vector.

 As mentioned above, we derive the spin dependent amplitude using a hybrid approach which is formulated in
 the covariant gauge.  The additional exchanged gluon is longitudinally polarized in such covariant
 gauge calculation. Unlike a quark scattering off the classical color field of  nucleus, the multiple scattering
 of this longitudinally polarized gluon with the background gluon field of nucleus can not be simply
 described by a Wilson line in the CGC formalism.
 Instead, the expression for  the gauge field created through the fusion of the
incoming longitudinally polarized gluon from the proton and small $x$ gluons from the nucleus takes a quite complicate form, and
contains  both singular terms (proportional to $\delta(x^+)$) and regular terms~\cite{Blaizot:2004wu},
\begin{eqnarray}
A^\mu(q)=A^\mu_{reg}(q)+\delta^{\mu-} A^-_{sing}(q) \ .
\end{eqnarray}
The regular terms $A^\mu_{reg}$ are given by
\begin{eqnarray}
A^\mu_{reg}&= & A_p^\mu \nonumber \\ &+& \frac{ig}{q^2+iq^+\epsilon}
\int \frac{d^2 p_\perp}{(2\pi)^2} \left \{ C_U^\mu(q,p_\perp) \left
[ \tilde U(k_\perp)-(2\pi)^2 \delta(k_\perp) \right ]  \right .\
\nonumber \\ && \ \ \ \ \ \ \ \ \ \ \ \ \ \  + \left .\
C^\mu_{V,reg}(q) \left [ \tilde V(k_\perp)-(2\pi)^2 \delta(k_\perp)
\right ] \right \} \frac{\rho_p(p_\perp)}{p_\perp^2} \label{dressed}
\end{eqnarray}
where $\rho_p(p_\perp)$ is the color source distribution inside a
proton,
 and $A_p^\mu$ is the gauge field created by the proton alone.
In the second term of the formula \ref{dressed}, $p_\perp$ is the
momentum carried by the incoming gluon from the proton and $k_\perp$
defined as $k_\perp=q_\perp-p_\perp$ is the momentum coming from the
nucleus. For the polarized case, there exists a correlation between
the transverse momentum $p_\perp$ and the transverse proton spin
vector $S_\perp$. Such a correlation is described
by the ETQS function, and leads to a SSA for direct photon
production~\cite{Schafer:2014zea} as well as Drell-Yan lepton pair production.
 The four vectors $C_U^\mu(q,p_\perp)$ and $C_{V,reg}^\mu$ are given by the following relations,
\begin{eqnarray}
&& C^+_U(q,p_\perp)=-\frac{{ p}_\perp^2}{q^-+i\epsilon}, \ \
C_U^-(q,p_\perp)=\frac{{k}_\perp^2-{q}_\perp^2}{q^++i\epsilon}, \ \
C^i_U(q,p_\perp)=-2{  \rm p}_\perp^i
 \\
 &&
C_{V,reg}^\mu(q)=2q^\mu-\delta^{-\mu} \frac{q^2}{q^++i\epsilon}
\end{eqnarray}
where the subscript $'reg'$ indicates that the corresponding term of
$A^\mu$ does not contain any $\delta(x^+)$ when expressed in
coordinate space. Here, we specified the $q^+$ pole structure
according to the fact that this term arises from an initial state
interaction. It is crucial to keep the imaginary part of this pole
in order to generate the non-vanishing spin asymmetry. The notation
${\rm p}_{\perp}$ is used to denote four dimension vector
 with  $p_{\perp}^2=-{\rm p}_\perp^2$.
$\tilde U(k_\perp)$ and $\tilde V(k_\perp)$ are the Fourier
transform of Wilson lines in the adjoint representation,
\begin{eqnarray}
\tilde U(k_\perp)&=&\int d^2 x_\perp e^{ik_\perp \cdot x_\perp}
{\cal P} {\rm exp} \left [ ig \int_{-\infty}^{+\infty} dz^+ A_A^-(z^+,x_\perp) \cdot T \right ] ,
 \\
\tilde V(k_\perp)&=&\int d^2 x_\perp e^{ik_\perp
\cdot x_\perp}
{\cal P} {\rm exp} \left [ i\frac{g}{2} \int_{-\infty}^{+\infty} dz^+ A_A^-(z^+,x_\perp) \cdot T \right ]
\end{eqnarray}
where the $T$ are the generators of the adjoint representation. The
singular terms reads,
\begin{eqnarray}
A^-_{sing}(q)=-\frac{ig}{q^++i\epsilon} \int \frac{d^2
p_\perp}{(2\pi)^2} \left [ \tilde V(k_\perp)-(2 \pi)^2
\delta(k_\perp) \right ] \frac{\rho_p(p_\perp)}{p_\perp^2}
\end{eqnarray}
The peculiar Wilson line $\tilde V$ differs from the normal one
$\tilde U$ by a factor $1/2$ in the exponent. It appears to be a generic feature that
all terms containing $\tilde V$ cancel eventually
when computing  a physical observable~\cite{Blaizot:2004wu,Blaizot:2004wv,Schafer:2014zea,Schafer:2014xpa}.

Following the method outlined in Ref.~\cite{Blaizot:2004wv}, one has to calculate the contributions from the regular terms
and the singular terms separately. In the prompt photon production case, the imaginary phase
necessary for non-vanishing SSAs is generated from the soft gluon pole£¬
while for Drell-Yan production, the imaginary phase arises from both the soft gluon pole and the hard gluon pole
due to the existence of an additional hard scale $M$.

The derivation of the spin dependent amplitude which contains the soft gluon pole contribution
is very similar to that presented in Ref.~\cite{Schafer:2014zea}.
The final expression for this amplitude takes form,
\begin{eqnarray}
{\cal M}_{SGP}&=&-ie g^2  \int \frac{d^2 p_\perp}{(2\pi)^2}
 \frac{\rho_{p,a}(p_\perp)}{p_\perp^2} \int \frac{d k^-_1 d^2 k_{1\perp} }{(2\pi)^3}
 \ \bar u(l_q)
    \nonumber \\
&& \times  \left \{
  \varepsilon \!\!\!/
S_F(x_1P+q)
 \frac{C_U \!\!\!\!\!\!\!/  \ \ (q-k_1,p_\perp)}{(q-k_1)^2+i\epsilon}  t^b S_F(x_1P+k_1) n\!\!\!/U(k_{1\perp})
 \right .\
 \nonumber \\ && \  \ \ \ \ +
  \frac{C_U \!\!\!\!\!\!\!/  \ \ (q-k_1,p_\perp)}{(q-k_1)^2+i\epsilon}  t^b S_F(x_1P-l_{\gamma^*}+k_1)
 n\!\!\!/ U(k_{1\perp})
S_F(x_1P-l_{\gamma^*}) \varepsilon \!\!\!/
 \nonumber \\ &&
\left .\  \ \ \ \  + \frac{C_U \!\!\!\!\!\!\!/  \ \
(q-k_1,p_\perp)}{(q-k_1)^2+i\epsilon}  t^b S_F(x_1P-l_{\gamma^*}+k_1)
\varepsilon \!\!\!/
   S_F(x_1P+k_1) n\!\!\!/ U(k_{1\perp})
\right \}
 \nonumber \\ && \times u(x_1P)
 \left [ \tilde U(k_\perp-k_{1\perp})-(2\pi)^2 \delta(k_\perp-k_{1\perp}) \right ]_{ba}
  \nonumber \\
&+& ieg^2 \int \frac{ d^2 p_\perp}{(2\pi)^2}
\frac{\rho_{p,a}(p_\perp)}{p_\perp^2} \int d^2 x_\perp e^{ik_\perp
\cdot x_\perp} \ \nonumber \\ & \times& \!\!\! \bar u(l_q) \frac{  n
\!\!\!/ S_F(x_1P-l_{\gamma^*}) \varepsilon \!\!\!/  + \varepsilon
\!\!\!/S_F(x_1P+q) n \!\!\!/ }{x_gP+i\epsilon} t^b
   U( x_\perp)  u(x_1P)\left [ \tilde U(x_\perp)-1 \right ]_{ba}
   \label{pamp}
\end{eqnarray}

The hard gluon pole is generated when the quark propagator
$S_F(x_1P-l_{\gamma^*})=i \frac{x_1P\!\!\!\!/-l_{\gamma^*}\!\!\!\!\!\!\!\!\!/}{(x_1P-l_{\gamma^*})^2+i\epsilon}$
 goes on shell.  It provides a phase proportional to $\delta(x_g-\bar x_g)$
where $x_g\equiv x-x_1=\bar x_g=x-M^2/2P\cdot l_{\gamma^*} $.
Diagrams Fig.1a-Fig.1d and Fig.1i-Fig.1l give rise to the
hard gluon pole contribution. The corresponding spin dependent amplitude reads,
\begin{eqnarray}
{\cal M}_{HGP}&=&-ie g^2  \int \frac{d^2 p_\perp}{(2\pi)^2}
 \frac{\rho_{p,a}(p_\perp)}{p_\perp^2} \int \frac{d k^-_1 d^2 k_{1\perp} }{(2\pi)^3}
 \ \bar u(l_q)
    \nonumber \\
&& \times  \left \{
    \frac{C_U \!\!\!\!\!\!\!/  \ \ (q-k_1,p_\perp)}{(q-k_1)^2+i\epsilon}  t^b S_F(x_1P-l_{\gamma^*}+k_1)
 n\!\!\!/ U(k_{1\perp})
S_F(x_1P-l_{\gamma^*}) \varepsilon \!\!\!/  \right .\
 \nonumber \\ &&
\left .\  \ \ \ \  + \frac{C_U \!\!\!\!\!\!\!/  \ \
(q-k_1,p_\perp)}{(q-k_1)^2+i\epsilon}  t^b S_F(x_1P-l_{\gamma^*}+k_1)
\varepsilon \!\!\!/
   S_F(x_1P+k_1) n\!\!\!/ U(k_{1\perp})
\right \}
 \nonumber \\ && \times u(x_1P)
 \left [ \tilde U(k_\perp-k_{1\perp})-(2\pi)^2 \delta(k_\perp-k_{1\perp}) \right ]_{ba}
  \nonumber \\
&+& \!\!  ieg^2  \!\!\! \int \frac{ d^2 p_\perp}{(2\pi)^2}
\frac{\rho_{p,a}(p_\perp)}{p_\perp^2} \int d^2 x_\perp e^{ik_\perp
\cdot x_\perp} \
 \nonumber \\& & \times \left \{ \bar u(l_q)
   \frac{  n
\!\!\!/ S_F(x_1P-l_{\gamma^*}) \varepsilon \!\!\!/ }{x_gP+i\epsilon} t^b
   U( x_\perp)  u(x_1P)\left [ \tilde U(x_\perp)-1 \right ]_{ba}
  \right .\
\nonumber \\ && \ \ \ \ \  +
i \bar u(l_q) p\!\!/ t^a S_F(x_1P-l_{\gamma^*}+k) n\!\!\!/ \left [ U(x_\perp)-1 \right]
S_F(x_1P-l_{\gamma^*}) \varepsilon \!\!\!/ u(x_1P)
\nonumber \\ &&   \ \ \ \ \ +
i  \bar u(l_q) n\!\!\!/ \left [ U(x_\perp)-1 \right]
S_F(l_q-k) p\!\!/ t^a S_F(x_1P-l_{\gamma^*})
 \varepsilon \!\!\!/ u(x_1P) \Big\}
 \label{hgp}
\end{eqnarray}
The last two terms in the above formula come from  the $A_p$ part of $A_{reg}$ in Fig.1b and Fig.1c.
Note that  Fig.1c contains no soft gluon pole,
while the soft gluon pole contribution from Fig.1b cancel out with its conjugate diagram.
With these two derived amplitudes, it is difficult to compute the full twist-3
 polarized cross section. However, in the kinematic regions where the collinear approach or the
 TMD factorization  approach is applicable, the calculation  can be greatly simplified, such that
 we can make comparisons among different formalisms.

\section{The SSA at low and high transverse momentum}
The SSA of Drell-Yan production can be described in the context of  the collinear higher twist factorization approach
at high transverse momentum $k_\perp^2 \sim Q_s^2 \ll l_{\gamma^*\perp}^2 $,  and  the TMD factorization framework
at low transverse momentum $l_{\gamma^*\perp}^2 \ll M^2$. In this section, we compare our hybrid approach with
these two formalisms in the corresponding kinematic regions.

We first extrapolate the full amplitudes to high transverse momentum region by Taylor expanding the hard parts
in terms of $k_{1\perp}$.
Repeating the power counting analysis in Ref.~\cite{Schafer:2014zea} and neglecting  power suppressed contributions,
the full amplitudes can be dramatically simplified to,
\begin{eqnarray}
{\cal M}_{SGP}&\approx&  - ieg^2 \int \frac{ d^2
p_\perp}{(2\pi)^2} \frac{\rho_{p,a}(p_\perp)}{p_\perp^2} \int d^2
x_\perp e^{ik_\perp \cdot x_\perp} \ \bar u(l_q)
 \nonumber \\ & &
\times
 \frac{  C_L \!\!\!\!\!\!\!/   \ \ (q,p_\perp)
S_F(x_1 P-l_{\gamma^*}) \varepsilon \!\!\!/  + \varepsilon \!\!\!/S_F(x_1 P+q)
C_L \!\!\!\!\!\!\!/ \ \ (q,p_\perp)  }{q^2+i\epsilon} t^b
 U( x_\perp)  u(x_1 P)\left [ \tilde U(x_\perp)-1 \right ]_{ba}
\nonumber \\
 {\cal M}_{HGP} &\approx&  - ieg^2 \int \frac{ d^2
p_\perp}{(2\pi)^2} \frac{\rho_{p,a}(p_\perp)}{p_\perp^2} \int d^2
x_\perp e^{ik_\perp \cdot x_\perp} \ \nonumber \\ & &
\times \left \{
 \bar u(l_q)  \frac{  C_L \!\!\!\!\!\!\!/   \ \ (q,p_\perp)
S_F(x_1 P-l_{\gamma^*}) \varepsilon \!\!\!/   }{q^2+i\epsilon} t^b
 U( x_\perp)  u(x_1 P)\left [ \tilde U(x_\perp)-1 \right ]_{ba} \right .\
\nonumber \\ && \ \ \ \ \  -
i \bar u(l_q) p\!\!/ t^a S_F(x_1P-l_{\gamma^*}+k) n\!\!\!/ \left [ U(x_\perp)-1 \right]
S_F(x_1P-l_{\gamma^*}) \varepsilon \!\!\!/ u(x_1P)
\nonumber \\ &&   \ \ \ \ \ -
i  \bar u(l_q) n\!\!\!/ \left [ U(x_\perp)-1 \right]
S_F(l_q-k) p\!\!/ t^a S_F(x_1P-l_{\gamma^*})
 \varepsilon \!\!\!/ u(x_1P) \Big \}
\end{eqnarray}
where $C_L \!\!\!\!\!\!\!/ \ \  $ is the well known effective Lipatov vertex for the production of a gluon
via the fusion of two gluons, and given by,
\begin{eqnarray}
\frac{C_L \!\!\!\!\!\!\!/}{q^2+i\epsilon}=\frac{C_U \!\!\!\!\!\!\!/}{q^2+i\epsilon}
-\frac{n\!\!\!/}{q^++i\epsilon}
\end{eqnarray}
The next step is to further expand the  hard part in terms of $k_\perp$ and keep the leading
power contribution following the method outlined in Section 3.3 in Ref.~\cite{Schafer:2014zea}.
After having done so, it becomes evident that the hard coefficient calculated from the above amplitude
is the same as that computed in the standard collinear twist-3 factorization~\cite{Ji:2006vf}.
On the other hand, using the Fierz identity,
the soft part from the nucleus side, i.e. Wilson lines, can be reorganized and expressed into
two parts: unpolarized gluon distribution and the novel gluon distribution $G_4$.
Collecting the hard gluon pole contribution and the soft gluon pole contribution,
 we eventually obtain the following polarized differential cross section,
\begin{eqnarray}
&& \!\!\!\!\!\!\!\!\!\!\!\!\!\!\!\!\!\!\!\!
 \frac{d\Delta \sigma}{ dM^2 d^2 l_{\gamma^*\perp}dy} = \frac{4 \pi \alpha_{em}^2}{3N_c  S M^2}
\frac{\alpha_s}{4 \pi^2}  \epsilon_\perp^{\alpha \beta} S_{\perp \alpha}l_{\gamma^*\perp \beta }
 \sum_q e_q^2 \int \frac{dx }{x} \frac{dx_g'}{x_g'}
\delta(\hat s+\hat t +\hat u -M^2) \frac{1}{- \hat u}
\nonumber \\ &\times&  \!\! \left \{ \left [
H_{Born}\left ( T_{F,q}(x,x) -x\frac{d}{dx} T_{F,q}(x,x) \right )+
N_s T_{F,q}(x,x) \right ]
  \left [  G_{DP}(x_g')-  G_4(x_g') \right ] \right .\
\nonumber \\ && \left .\ + \
 T_{F,q}(x-\bar x_g,x)
 \Big [ H_{HGP}^a \left  [  G_{DP}(x_g')-  G_4(x_g') \right ]
 +H_{HGP}^b   G_{DP}(x_g') \Big ] \right \}
 \label{cross}
\end{eqnarray}
where the hard coefficients are given by,
\begin{eqnarray}
H_{Born}&=&\frac{N_c^2}{N_c^2-1} \left [ - \frac{\hat s}{\hat t}-\frac{\hat t}{\hat s}-\frac{2M^2\hat u}{\hat s \hat t} \right ]
\\
N_s&=&\frac{N_c^2}{N_c^2-1} \frac{M^2}{\hat s \hat t^2} \left [M^4-2M^2 \hat t+\hat u^2 \right ]
\\
H_{HGP}^a&=&-\frac{N_c^2}{N_c^2-1} \frac{(M^2-\hat t)^3+M^2 \hat u^2}{\hat s \hat t^2}
\\
H_{HGP}^b&=&\frac{\hat s}{\hat s +\hat u} \frac{(M^2-\hat t)^3+M^2 \hat u^2}{\hat s \hat t^2}
\end{eqnarray}
with $\hat s, \hat u, \hat t$ being the normal partonic Mandelstam variables. In the collinear limit
 $l_{\gamma^*\perp}^2\gg k_\perp^2$, they can be expressed as,
\begin{eqnarray}
\hat s= \frac{l_{\gamma^*\perp}^2}{(1-z)(1-\tilde z)}
\ \ \ \ \ \  \hat u= -\frac{l_{\gamma^*\perp}^2}{1-z}
\ \ \ \ \ \ \hat t=-\frac{l_{\gamma^*\perp}^2}{1-\tilde z}
\end{eqnarray}
where $\tilde z$ is the longitudinal momentum fraction of the incoming gluons $x_g' \bar P$ carried by the virtual photon.
In the above formula, $x_g' G_{DP}(x_g')$ and $x_g' G_4(x_g',k_\perp)$
 are the integrated gluon distributions defined as
$x_g' G_{DP}(x_g')=\int d^2k_\perp x_g' G_{DP}(x_g',k_\perp)$,
 $x_g' G_4(x_g')=\int d^2k_\perp  x_g' G_4(x_g',k_\perp)$, respectively.
 The gluon distribution $G_4(x_g',k_\perp)$  possesses an unique Wilson line structure,
\begin{eqnarray}
x_g' G_{4}(x_g', k_\perp)&=& \frac{k_\perp^2 N_c}{2 \pi^2 \alpha_s}
\int \frac{d^2x_\perp d^2y_\perp }{(2\pi)^2} e^{i  k_\perp \cdot
(x_\perp-y_\perp)} \frac{1}{N_c^2} \langle {\rm Tr_c} [U(x_\perp)]
{\rm Tr_c}[ U^\dag(y_\perp)] \rangle_{x_g'} \label{g4}
\end{eqnarray}
It has been shown in the MV model~\cite{McLerran:1993ni} that $G_4$ is suppressed by the power of $1/N_c^2$  as
compared to the unpolarized dipole distribution $G_{DP}$~\cite{Schafer:2014zea}.
If we neglect these  $1/N_c^2$ suppressed terms that essentially arises from color entanglement effect,
 it is easy to see that the polarized cross section computed in the standard collinear twist-3
 approach can be recovered from our hybrid approach.

To compare with the result from TMD factorization, we have to further extrapolate the above result to the
moderate transverse momentum region, $Q_s^2<< l_{\gamma^* \perp}^2<< M^2$ where TMD factorization is applicable.
 This can be easily done by expanding the delta function,
 \begin{eqnarray}
\delta(\hat s+\hat t +\hat u -M^2)=\frac{1}{\hat s}
\left [ \frac{\delta(1-z)}{(1-\tilde z)_+} + \frac{\delta(1-\tilde z)}{(1- z)_+}
+\delta(1-z)\delta(1-\tilde z) {\rm ln}\frac{M^2}{l_{\gamma^* \perp}^2} \right ]
 \end{eqnarray}
 where only the first term proportional
to $\delta(1-z)$ gives rise to the leading power contribution. In the limit $z\rightarrow 1$, one has
$|\hat t|<<|\hat u| \sim \hat s$, which implies $\bar x_g\rightarrow 0$. The fact that the hard gluon
pole degenerates with the soft gluon pole in this kinematic region allows us to combine two contributions together.
Substituting the above expansion into Eq.~\ref{cross}, one ends up with,
\begin{eqnarray}
&& \frac{d\Delta \sigma}{ dM^2 d^2 l_{\gamma^*\perp}dy}  = \frac{4 \pi \alpha_{em}^2}{3N_c S M^2}
\frac{\alpha_s}{4 \pi^2}  \frac{\epsilon_\perp^{\alpha \beta} S_{\perp \alpha}l_{\gamma^*\perp \beta }}{l_{\gamma^*\perp}^4}
\nonumber \\&& \ \ \ \ \ \  \times
 \sum_q e_q^2 \int \frac{dx }{x} \frac{dx_g'}{x_g'}
\delta(1-z)  T_{F,q}(x,x)
  G_{DP}(x_g') \left [ \tilde z^2+(1-\tilde z)^2 \right ]
 \label{2}
\end{eqnarray}
which agrees with Eq.31 in Ref.~\cite{Ji:2006vf}. As expected, all terms arising from the color entanglement
effect cancel out in the kinematic limit we consider. Therefore, our hybrid approach is consistent with
TMD factorization at moderate transverse momentum.

However, TMD factorization applies in a broader kinematic region: $ l_{\gamma^*\perp}^2 \ll M^2$.
On the other hand, the hybrid approach is valid as long as $\Lambda_{QCD}^2 \ll l_{\gamma^*\perp}^2$.
To demonstrate the complete equivalence between the two formalisms in the overlap region $\Lambda_{QCD}^2 \ll l_{\gamma^*\perp}^2\ll M^2$,
 it is necessary to keep $k_\perp$ finite when computing the hard coefficients.
This makes the evaluation of the polarized cross section  much more involved.
Nevertheless, through the explicit calculation, we verify that the SSA does not receive the contribution
from the initial state interaction due to the complete cancelation between the soft gluon pole contribution and
the hard gluon pole contribution in the low transverse momentum region.
We are thus  left with  the hard gluon pole contribution from diagrams Fig.1b and Fig.1c. The corresponding
amplitudes are give by the last two terms in Eq.~\ref{hgp}. We further found that the final state interaction
shown in Fig.1b only  yields the power suppressed contribution at low transverse momentum.
The only remaining piece is the hard gluon pole contribution from diagram Fig.1c
which can be computed following the standard procedure. The fact that the entire surviving contribution
is just due to hard pole at low transverse momentum has also been observed in Ref.~\cite{Ratcliffe:2007ye}.
 At this step, we would like to mention that the soft part associated with the diagram Fig.1c only
 contains the regular Wilson line structure.
In the end, the polarized cross section at low transverse momentum can be nicely cast into the following compact form,
\begin{eqnarray}
&& \frac{d\Delta \sigma}{ dM^2 d^2 l_{\gamma^*\perp}dy}  =
\frac{4\pi \alpha_{em}^2}{3N_cS M^2}\frac{\alpha_s}{4 \pi^2}\frac{1}{2}
 \nonumber \\&& \ \ \ \ \ \  \times
 \sum_q e_q^2
 T_{F,q}(\xi,\xi)    \int \frac{ d x_g'}{x_g'}
\int  \frac{ d^2 k_\perp}{k_\perp^2} G_{DP}(x_g',k_\perp)
\epsilon_\perp^{\alpha \beta} S_{\perp \alpha}
\frac{-\partial A(k_\perp,l_{\gamma^*\perp}, \tilde z)}{\partial l_{\gamma^*\perp}^\beta}
 \label{3}
\end{eqnarray}
with $A(k_\perp,l_{\gamma^*\perp}, \tilde z)$ being defined as,
\begin{eqnarray}
 A(k_\perp,l_{\gamma^*\perp}, \tilde z)=\left [
 \frac{ l_{\gamma^*\perp} |l_{\gamma^*\perp}-k_\perp|}{(1-\tilde z) l_{\gamma^*\perp}^2+\tilde z(l_{\gamma^*\perp}-k_\perp)^2}
 -\frac{l_{\gamma^*\perp}-k_\perp}{|l_{\gamma^*\perp}-k_\perp|} \right ]^2
 \label{6}
\end{eqnarray}
The transverse momentum carried by small $x$ gluons is of the order of the saturation scale $Q_s$.
At moderate transverse momentum  $M^2 \gg l_{\gamma^*\perp} \gg Q_s\sim k_\perp$,
 one can Taylor expand the hard coefficient
 $\partial A(k_\perp,l_{\gamma^*\perp}, \tilde z)/\partial l_{\gamma^*\perp}^\beta$ in terms of $k_\perp$.
 By keeping the nontrivial leading term and neglecting the terms suppressed by the power of $k_\perp/ l_{\gamma^*\perp}$,
   Eq.~\ref{2} can be readily recovered   from Eq.~\ref{3}.

As mentioned above, the SSA for the Drell-Yan production at low transverse momentum also can be described in the
TMD factorization approach.
 The corresponding polarized cross section reads,
\begin{eqnarray}
&& \frac{d\Delta \sigma}{ dM^2 d^2 l_{\gamma^*\perp}dy}  =
\frac{4 \pi \alpha_{em}^2}{3N_c  SM^2}
 \nonumber \\&& \ \ \ \ \ \  \times
 \sum_q e_q^2 \int d^2 p_\perp d^2 p_\perp'
 \frac{\epsilon_\perp^{\alpha \beta} S_{\perp \alpha}p_{\perp \beta} }{M_p}
 \delta^2(l_{\gamma^*\perp}-p_\perp-p_\perp')
 f_{1T,q}^\perp(x,p_\perp) \bar f_q(x', p_\perp')
 \label{4}
\end{eqnarray}
where $f_{1T,q}^\perp(x,p_\perp)$  and $\bar f_q(x', p_\perp')$ are the quark Sivers function  and the
unpolarized anti-quark distribution from the target nucleus, respectively.
At small $x$, the anti-quark  distribution is dynamically generated through gluon splitting process~\cite{McLerran:1998nk},
\begin{eqnarray}
\bar f_q(x', p_\perp')=\frac{\alpha_s}{4 \pi^2} \int \frac{d x_g'}{x_g'}
\int  \frac{ d^2 k_\perp}{k_\perp^2} G_{DP}(x_g',k_\perp) A(k_\perp,p_\perp', \tilde z)
\label{5}
\end{eqnarray}
As argued above, the typical small $x$ anti-quark transverse momentum $p_\perp'=l_{\gamma^*\perp}-p_\perp$ is of the order of $Q_s$
 and much larger than the incoming quark transverse momentum $p_\perp\sim \Lambda_{QCD}$.
 After substituting Eq.~\ref{5} into Eq.~\ref{4} and carrying out the integration over $p_\perp'$,
 we make Taylor expansion for $ A(k_\perp,l_{\gamma^*\perp}-p_\perp, \tilde z)$ in terms of $p_\perp$
 and keep the linear term.
To make the connection between Eq.~\ref{4} and Eq.~\ref{3},
one further use the well known relation $T_{F,q}=\int d^2 p_\perp \frac{p_\perp^2}{M_p} f_{1T,q}^\perp(x,p_\perp)$ for
the Drell-Yan process~\cite{Boer:2003cm}.
 It is then straightforward to reproduce Eq.~\ref{3} from Eq.~\ref{4}.

 We thus confirm that the hybrid approach agrees with TMD factorization for the Drell-Yan production
 in the full overlap kinematical region where both  the formalisms are  applicable. This agreement provides
 strong evidence that the hybrid approach is complete and self-consistent.
 We  notice that the equivalence  between the small $x$ formalism and the TMD factorization approach in describing
  the same observable has  also been verified in Ref.~\cite{Kang:2012vm}.

\section{Conclusion}
Color entanglement effect is usually believed to be absent in the Drell-Yan process in the context of TMD
factorization because of the simple color flow structure, though a complete consensus has not yet been reached~\cite{Buffing:2013dxa}.
 With the help of  newly developed hybrid approach, we are able
for the first time to provide a non-trivial check on this statement for the SSA case.
To be more precise, we take into account one additional gluon exchange from polarized proton and resum gluon
scattering to all orders on nucleus side using the hybrid approach. It has been shown that
 in the polarized cross section, all terms arising
from the color entanglement effect drop out at low transverse momentum due to the systematical cancelation
between the soft gluon pole contribution and the hard gluon pole contribution. However, at high transverse momentum,
 the polarized cross section computed in the hybrid approach differs from that obtained from the standard collinear
 twist-3 approach by some additional contributions whose emergence can be attributed to color entanglement.
 Such novel color entanglement effect in principle could be studied at RHIC~\cite{Aschenauer:2013woa},
 though it is found to be $1/N_c^2$ suppressed and thus very small.

\vskip 1 cm {\bf Acknowledgments:}
 I would like to thank Daniel Boer for interesting conversations about some conceptual issues.
  This research has been supported by BMBF (OR 06RY9191), and the EU "Ideas" program QWORK (contract 320389).

\end {document}